\documentclass[conference]{IEEEtran}
\IEEEoverridecommandlockouts
\usepackage{cite}
\usepackage{amsmath,amssymb,amsfonts}
\usepackage{graphicx}
\usepackage{textcomp}
\usepackage{xcolor}
\usepackage{booktabs}
\usepackage{tabularx}
\usepackage{algorithm}
\usepackage{algorithmic}
\usepackage{cleveref}
\usepackage{array}
\usepackage{enumitem} 
\newtheorem{definition}{Definition}[section]
\newcolumntype{L}[1]{>{\raggedright\arraybackslash}p{#1}}
\newcolumntype{Y}{>{\raggedright\arraybackslash}X}
\newcolumntype{L}[1]{>{\raggedright\arraybackslash}p{#1}}
\newcolumntype{Y}{>{\raggedright\arraybackslash}X}
\def\BibTeX{{\rm B\kern-.05em{\sc i\kern-.025em b}\kern-.08em
    T\kern-.1667em\lower.7ex\hbox{E}\kern-.125emX}}
\begin{document}

\title{Interference-Aware Cross-Application Placement: A Multi-Objective Optimization Approach for Microservice Clusters}
\author{\IEEEauthorblockN{1\textsuperscript{st} Iqra Zafar}  
\IEEEauthorblockA{ 
\textit{Hasso Plattner Institute}\\
Potsdam, Germany\\
iqra.zafar@hpi.de}
\and
\IEEEauthorblockN{2\textsuperscript{nd} Christian Medeiros Adriano}
\IEEEauthorblockA{
\textit{Hasso Plattner Institute}\\
Potsdam, Germany \\
christian.adriano@hpi.de}
\and
\IEEEauthorblockN{3\textsuperscript{rd} Holger Giese}
\IEEEauthorblockA{
\textit{Hasso Plattner Institute}\\
Potsdam, Germany \\
holger.giese@hpi.de}
}

\maketitle

\begin{abstract}
In modern cloud architectures, multiple applications often run within the same clustered environment, sharing underlying resources. This resource sharing can cause interference among applications, leading to degraded latency and reduced system stability. As containerized microservices become increasingly central to cloud-native applications, their performance can suffer from complex interference scenarios related to resource competition. Meanwhile, most existing microservice approaches address interference either by detecting and localizing performance issues or by optimizing latency alone, without explaining why specific co-locations cause cross-application interference, and how this can inform service placement optimization. This work closes that gap by building a spatio-temporal data structure that captures the causal effects of cross-application interference. These causal effects are mathematically formalized as necessary and sufficient conditional probabilities that inform a multi-objective optimizer (Optuna). Cross-application profiling is used to simulate traces and estimate interference probabilities, while per-service latency baselines are provided by performance data, such as 95th-percentile response times (p95). Our approach supports network penalties, application isolation requirements, and adjustable weighting of necessary and sufficient causal metrics. Experimental results on real multi-application workloads show that interference-aware placements significantly reduce cross-application interference and improve response performance. Ultimately, the causality-driven multi-objective formulation gives cloud operators explicit control over interference, latency, and communication overhead when configuring service placements.
\end{abstract}


\section{Introduction}
Microservices have become the foundation of scalable cloud-native systems (e.g., in Kubernetes and serverless environments) \cite{newman2021building}. A monolithic program may be divided into smaller, independently deployable, and scalable services, i.e., microservices. Despite its adaptability, this architecture exhibits unpredictable performance due to co-location interference (\Cref{def:interference}) over shared resources. Inadequate service placement can cause latency to rise, response times to deteriorate, or even breach of SLAs and SLOs targets.
Service interference is a common cause of performance degradation in current cloud-native environments, often leading to latency spikes, dropped requests, or reduced throughput~\cite{Adeppady2022iPlace,zafar2024towards}. Co-located microservices competing for constrained CPU, memory, or I/O bandwidth on shared nodes are usually the source of this interference.

\begin{definition}\label{def:interference} \textbf{Interference} happens when two services that have no logical dependency (caller-callee relation) but compete for the same resource (compute, memory, I/O) to the extent that they affect each other's performance (e.g., throughput, latency)~\cite{pu2010interference}.
\end{definition}

Traditional placement techniques lack the foresight to proactively prevent such interference, as they often rely on reactive~\cite{Adeppady2022reducing,Souza2020IRENE,Song2025CKoordinator} or static/periodic optimization~\cite{Qu2024CommunityPlacement}. The reliance on purely correlational information and static rules hinders current approaches' ability to identify the origin of cross-application interference and the direction of its effects, given the complex propagation of performance anomalies. To move beyond reacting to performance breaches after they occur and searching for elusive sources across multiple applications, we propose capturing causal-effect associations and pathways that are not typically visible in microservice call graphs (i.e., service meshes).


\textbf{Research Questions:} 

\begin{itemize}
    \item \textbf{RQ.1 (Severity of phenomenon)}: \textit{What is the magnitude of the cross-application interference?} - Magnitude consists of the number of nodes and the level of performance degradation, which determines the phenomenon severity and consequently, the specific needs for mitigation, e.g., which nodes and the extent of interference reduction.
    \item \textbf{RQ.2 (Effectiveness)}: \textit{Is our service placement approach able to reduce interference across scenarios of increasing complexity (number of nodes affected)?} - This question evaluates the effectiveness of the approach in recovering the microservice systems from degradation, while also providing insights into the scope of adaptations, i.e., the number of impacted nodes and residual interference (unmitigated). For instance, are there trade-offs between the residual interference and impacted nodes? 
   \item \textbf{RQ.3 (Component contribution)}: \textit{What is the impact of each component of the approach on the effectiveness of placement?} - The components of the approach consist of two elements of causal knowledge that will be introduced later.
    
\end{itemize}

\textbf{Our contribution: } This paper presents a novel approach to microservice placement optimization by integrating formal causal reasoning with continuous adaptation within an offline feedback loop extension of the MAPE-K architectural pattern~\cite{krupitzer2015survey}. The main contributions are as follows:
\textbf{(1) A causality-aware microservice placement framework} -- We propose an adaptive placement approach based on Spatio-Temporal Interference Graphs (STIGs) from our prior work \cite{zafar2024stigs}. The framework uses Probability of Sufficiency (PS), Probability of Necessity and Sufficiency (PNS), and Probability of Necessity (PN) to identify, explain, and mitigate interference among co-located services. By grounding these metrics in formal notions of necessity and sufficiency, the approach improves both the precision and interpretability of placement decisions.
\textbf{(2) Runtime-aware adaptation within an offline MAPE-K loop} -- We integrate Prometheus and Istio to continuously collect traces, service- and system-level metrics, including response time, CPU/memory usage, latency, and error rates. These observations feed the Monitor and Analyze phases of the MAPE-K loop, enabling the system to detect SLO violations and performance anomalies, re-evaluate current placements through the STIG model, and respond to changing workload conditions.
\textbf{(3) A multi-objective adaptive planner for scalable placement optimization} -- We develop an Optuna-based planner that jointly optimizes service responsiveness and interference reduction. To keep the search efficient and focused, the planner prioritizes services using PN- and PS-based rankings, progressively expands the placement search space, and incorporates SLO-aware service importance during optimization.
\textbf{(4) A comprehensive evaluation on representative microservice benchmarks} -- We evaluate the proposed approach on Tea-Store, BookInfo, and Sock-Shop across diverse runtime scenarios, including failure conditions, traffic surges, and geo-distributed deployments. The results show that the approach consistently reduces interference, improves response performance, and adapts placement strategies in response to observed system dynamics. Finally, all data and scripts are publicly available\footnote{Reproducibility package: https://doi.org/10.5281/zenodo.18900711}.

 \noindent\textbf{Tour} - ~\Cref{sec:preliminaries} details the preliminaries,  \Cref{sec:sota} the state-of-the-art, \Cref{sec:approach} the service placement approach, \Cref{sec:evaluation} the evaluation, while implications and threats to validity are in \Cref{sec:discussion} and we conclude with \Cref{sec:conclusion}. \\

\section{Preliminaries}\label{sec:preliminaries}

\subsection{Contention and Interference}
\textbf{Contention} refers to the pressure a service exerts on a shared resource (e.g., CPU, memory, I/O bandwidth) when multiple services attempt to access it simultaneously~\cite{zhuravlev2010addressing}. High contention arises when demand for a resource approaches or exceeds its available capacity. This leads to performance degradation, e.g., increased latency or response time. In microservice systems, contention is a key driver of interference, as two services without logical dependencies (no caller–callee relation) may still negatively impact their performance when they contend for the same underlying resource~\cite{Adeppady2022iPlace}.

\textbf{Interference} arises when resource competition and degrade microservice performance \cite{Adeppady2022reducing}. We capture cross-application interference using a dynamic data structure (\Cref{def:STIG}), where dependencies between microservices are measured as correlations observed during concurrent executions. For more details about how STIGs are being generated from the system's knowledge graph, refer to our prior work~\cite{zafar2024stigs}.

\begin{definition}\label{def:STIG} \textbf{Spatio-Temporal Interference Graph (\textit{STIG})} is denoted as $\mathcal{G}=(V,E,X_{v(t)},X_{e(t)})$, where $V$ are nodes representing services, $E$ are directed edges representing interference between services across applications, $X_{v(t)}$ are the time-varying node features (e.g., resource per service), and $X_{e(t)}$ the edge features (e.g., interference probability).
\end{definition}

\subsection{Causality in STIG} \label{causility-STIG}

Microservice interference is inherently a causal phenomenon, where the resource consumption of one service affects the performance of other services. However, correlations in execution traces do not imply causation. To remediate that, we rely on causal reasoning about interference represented in STIGs and we adopt the counterfactual notions\footnote{For a formal analysis of 12 definitions of sufficiency and necessity, their use cases, and implications, please refer to Becker~\cite{beckers2021causal}.} of \textit{sufficiency} and \textit{necessity} from Pearl \cite{pearl2009causality} and Halpern \& Hitchcock \cite{halpern2010actual}. 

\begin{definition} \textbf{Causal Sufficiency in STIGs}\label{def:causal-sufficiency} A cause is sufficient for an effect if its occurrence alone guarantees the effect. Within a STIG, this corresponds to an interference edge $e \in E$, where the source service’s consumption always induces a performance degradation in the target service. A high-probability edge weight (close to 1.0) in an STIG can be interpreted as approaching sufficiency, meaning that whenever the source service experiences contention, the target service is consistently affected by interference. An example is when a service $A$ interferes with service $B$ with probability 1.0. This reflects a condition of near-sufficiency, because the simultaneous execution of service $A$ causally induces a slowdown in service $B$. Probability of sufficiency is defined in \Cref{eq:ps_algo}.
\begin{equation}
    \mathrm{PS}(X \rightarrow Y) =
    \begin{cases}
        \dfrac{\mathrm{PNS}(X \rightarrow Y)}{1 - P(Y \mid \neg X)}, &
            P(Y \mid \neg X) < 1, \\[0.5em]
        0, & \text{otherwise},
    \end{cases}
    \label{eq:ps_algo}
\end{equation}
\end{definition}

\begin{definition}\textbf{Causal Necessity in STIGs}\label{def:causal-necessity} A cause is necessary for an effect if the effect cannot occur in its absence. In STIG terms, this implies that interference observed at a target service node cannot be explained without a particular source node being active in the graph. Necessity can be assessed by analyzing minimal cut-sets or exclusive dependency paths in the STIG \cite{eiter2002causes, halpern2005causes}. If removing a given source node (or its edges) eliminates all interference paths leading to a target, that source is necessary. The obvious case is when Service $A$ is the only origin of interference paths leading to Service $B$, then $A$’s contention is necessary for $B$’s slowdown. Pearl's probability of necessity is defined in \Cref{eq:pn_algo}.
\begin{equation}
    \mathrm{PN}(X \rightarrow Y) =
    \begin{cases}
        \dfrac{\mathrm{PNS}(X \rightarrow Y)}{P(Y \mid X)}, &
            P(Y \mid X) > 0, \\[0.5em]
        0, & \text{otherwise}
    \end{cases}
    \label{eq:pn_algo}
\end{equation}
\end{definition}

\textbf{Interplay of Sufficiency and Necessity} -- Because of the large number of interdependencies in multi-tenant systems, we expect few interference relations are both necessary and sufficient. \textbf{Sufficiency} as a probability, it strongly increases the likelihood of interference, even though it does not guarantee it in every case. \textbf{Necessity} is distributed, because it is part of a set of causes connected by conjunctions, i.e., all must be present to cause an interference. In our formulation, we map sufficiency to edge weights and necessity to graph connectivity properties. This way the STIG can provide a causal lens for inferring direction in the correlations computed from execution traces and enable the identification of the interference propagation path across the microservice deployment. The probability of necessity and sufficiency is then computed as in \Cref{eq:pns_algo}.
\begin{equation}
    \mathrm{PNS}(X \rightarrow Y)
    = \max \bigl( 0,\,
        P(Y \mid X) - P(Y \mid \neg X)
      \bigr)
    \label{eq:pns_algo}
\end{equation}


\subsection{MAPE-K as the Control Framework}
Our system is designed around the MAPE-K (Monitor–Analyze–Plan–Execute over Knowledge) loop, a popular autonomic control architecture to allow self-adaptive behavior in microservice-based cloud settings~\cite{krupitzer2015survey}. This loop is offline, meaning it is not embedded or running continuously at runtime within the cloud system. The goal of MAPE-K is to systematically close the feedback loop between dynamic decision-making and system observation. For practical examples of how the MAPE-K can be applied to causal-based analysis and planning, please refer to our previous work~\cite{adriano2025neuro,zafar2024towards}.
\section{State-of-the-Art}\label{sec:sota}

In software architecture, many approaches apply multi-optimization models to improve system performance~\cite{aleti2012software}, inclusive distributed architectures. Concerning adaption of microservices, most state-of-the-art focus on detection-centric~\cite{soldani2022anomaly} and correlation-driven~\cite{gan2023sleuth} to surface latency symptoms (p95/p99) and localize "likely" culprits along a trace graph. However, these approaches often overlook \emph{why} a configuration produces interference and \emph{how} to adjust placement under competing objectives, such as increasing response time or mitigating interference.

\textbf{Interference‑aware placement} has been addressed by iPlace~\cite{Adeppady2022iPlace,Adeppady2022reducing}, which clusters microservices to minimize interference and reduce the number of servers while still meeting performance targets. IRENE~\cite{Souza2020IRENE} targets edge environments and uses a genetic algorithm to jointly optimize high availability and contention‑induced performance interference for microservice applications. Community- and priority-based microservice placement~\cite{Qu2024CommunityPlacement} models dependencies as a graph and incorporates long-term interference awareness into the placement optimization problem. At production scale, C‑Koordinator~\cite{Song2025CKoordinator} extends Kubernetes/Koordinator with CPI‑based prediction models to detect and mitigate interference in co‑located microservice clusters, improving response‑time metrics across.

\textbf{Optimal placement} of microservices has been investigated in various methods. Particle Swarm Optimization‑based microservice placement~\cite{Benamor2024PSOMicroservices}  allocates microservices in cloud‑native infrastructures, including a Sock Shop case study that demonstrates improved resource utilization and SLA compliance. Multi‑objective reconfiguration of real‑time microservices~\cite{Abadeh2025NROMicroservices} applies an energy‑based nuclear reaction optimization meta‑heuristic to adapt microservice deployments to multiple objectives, including timing and resource constraints. An Optuna-based~\cite{Akiba2019Optuna} multi‑objective optimization was adopted to fine-tune edge service placement and related parameters to minimize both latency and deployment cost for location-based services in mobile edge computing environments \cite{kittikamron2023edge}.

\textbf{Causal latency modeling} for microservices employs causal discovery and domain knowledge methods to build latency graphs that explain SLO violations and support prediction and diagnosis~\cite{lohse2025causal}. CAR‑PT and related causal performance testing approaches employ causal reasoning to generate workloads that are more likely to expose performance issues, linking causal graphs with performance‑critical test inputs~\cite{mascia2025callmit}. Detecting causal structure on cloud applications with Granger causality infers dependency graphs among microservices from logs and traces, providing causal structures that can complement or motivate representations such as STIGs~\cite{wang2021granger}. A recent survey of causal‑inference‑based RCA for microservices evaluates multiple causal discovery and reasoning techniques on benchmark datasets, and positions causal graphs as a foundation for moving from root‑cause localization toward broader decision problems such as placement~\cite{pham2024causalrca}.

\textbf{Methodological challenges} -- 
(i) Causality vs.\ correlation: since counterfactuals (for example, “what if these two services were not co-located?”) are typically missing~\cite{wang2024comprehensive,panahandeh2024serviceanomaly,soldani2022anomaly,lohse2025causal}, necessity and sufficiency remain indistinct, and methods overfit observed traces. (ii) Actuation granularity: although placement is a key trigger for interference, placement-aware actions (affinity/anti-affinity, node isolation) are seldom explored; remediation usually relies on auto-scaling, restarting, or circuit breaking~\cite{rajagopalan2015app}. (iii) Single-objective bias: many approaches optimize tail delay only and do not quantify the trade-off surface or weight interference reduction against response time or throughput~\cite{wu2020microrca}. (iv) Blind spots in evaluation: studies often under-stress behavioral complexity (spikes, node failures, location changes) and rarely separate it from structural complexity (graph size or density), while favoring steady-state traffic and small homogeneous topologies \cite{yao2024chain}. (v) Observability limitations: sensitivity to incomplete knowledge is rarely assessed; trace sampling, head/tail bias, and partial coverage are common, and few systems maintain a persistent knowledge base for design-time evaluation or reuse of placements~\cite{yao2024sparserca,zhang2022crisp}; finally, misreported overheads of online inference and continuous tracing reduce external validity and repeatability~\cite{allam2024synthetic}.
Our contribution is to bridge the detection-to-placement decision that is informed by causal knowledge and, hence, more robust to hidden changes in the architecture.

\section{Approach}\label{sec:approach}
\subsection{System Architecture \& Methodology}
\begin{figure}[htbp]
\centerline{\includegraphics[width=0.4\textwidth]{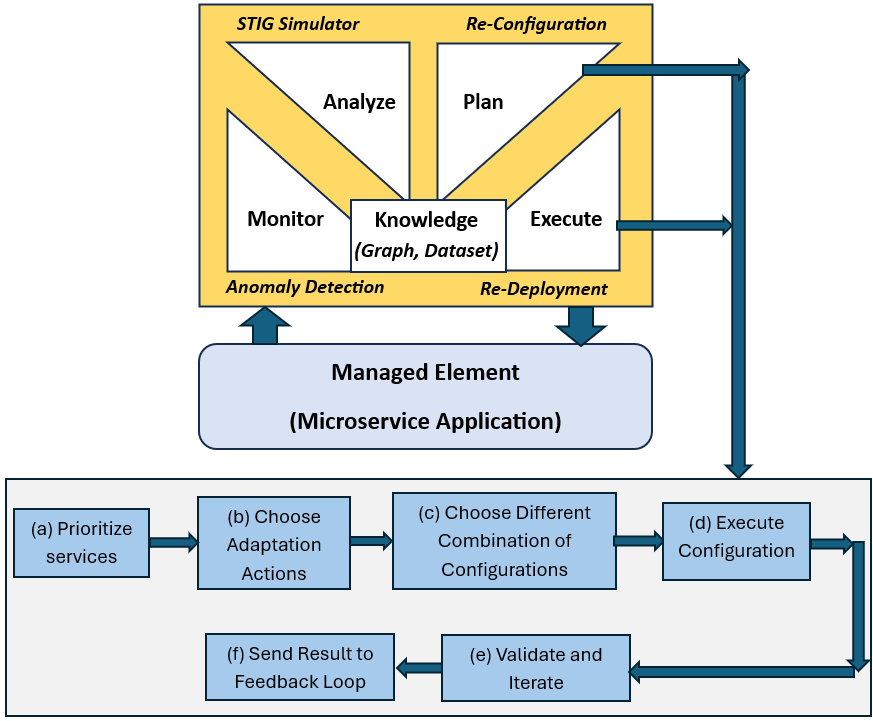}}
\caption{STIG-based MAPE-K loop with interference-aware service placement.}
\label{fig1}
\end{figure}

\Cref{fig1} shows the proposed self-adaptive loop, which follows the offline MAPE-K pattern. The managed element is a multi-tenant microservice application on Kubernetes. The controller detects interference between co-located services and proactively reconfigures their placement to improve performance and fault tolerance without human intervention.

In the \textbf{Monitor phase}, the system continuously collects runtime data: request traces, response-time statistics (for example p95 latency per service), resource usage (CPU, memory, I/O), deployment state (service replica to node mapping), and cross-application interactions. These metrics and traces are normalised and stored in the knowledge base as time-stamped events and call graphs.

In the \textbf{Analyze phase}, the STIG simulator derives interference knowledge from these observations. It builds STIG graphs that capture which services interfere when co-located in a same cluster sharing same host/server, and estimates causal measures (PN, PS, PNS) for each service pair. Combining these with STIG edge weights yields an expected interference score per pair. The analyzer then correlates these scores with response-time degradation to identify interference-prone services, producing STIGs, causal estimates, and per-service interference sensitivity indicators.

In the \textbf{Plan phase}, the controller computes a placement strategy. It aggregates pairwise interference scores into a per-service priority ranking, then defines a search space of alternative placements for the most interference-sensitive services. An Optuna-based multi-objective optimizer used, evaluating each candidate placement by total expected interference and response-time cost (p95 latency plus penalties for cross-node communication on critical edges). From the resulting Pareto front, the planner selects a configuration that matches current operator preferences and forwards it to Execute.

In the \textbf{Execute phase}, the system enacts the selected placement by updating Kubernetes scheduling constraints (i.e. node selectors) and allowing the orchestrator to migrate pods. It may move only high-priority services or rebalance all services. Monitoring then continues under live workload, allowing direct observation of the impact on latency, throughput, and interference indicators.

The shared \textbf{Knowledge} base stores deployment graphs, STIGs, causal interference estimates, service rankings, and outcomes of past reconfigurations (for example, “separating A from B reduced interference by X but increased latency by Y ms”). This history enables continuous refinement. The \textit{Analyze} phase updates interference models. The \textit{Plan} phase favours successful patterns, while the \textit{Execute} collects further traces for future analysis. Over time, the MAPE-K loop evolves from purely reactive control to experience-driven, interference-aware placement in a multi-node cluster.

\subsection{Algorithm} \label{algo}
Our research utilizes two algorithms,  \Cref{alg:prob} is used to compute causal probabilities values (PN,PS,PNS) from collected traces across multi-node cluster.The algorithm starts by loading the cleaned trace dataset $D$ from CSV (line 1). For each span, it parses the start time into a timestamp $t_s$ (line~-2),
converts the duration into seconds, and computes the end time $t_e = t_s + \text{duration}$ (line 3). For each service $s$, it computes the mean $\mu_s$ and standard deviation $\sigma_s$ of span durations (line~4),
and defines a service-specific anomaly threshold (line 5). A span is marked anomalous if it reports an error or if its duration
exceeds $\tau_s$ (line 6), forming the anomalous set $D_a$ (line 7). The algorithm then counts anomalous spans per service and retains only services with at least $M$ anomalies to reduce noise (line 8). For each remaining
service, anomalous spans are sorted by start time and stored as aligned arrays $\{t_s^x\}$ and $\{t_e^x\}$ (line 9). The output table $T$ is initialized next (line 10).
The function \texttt{TemporalHits} (lines 11-17) iterates over all events of service $X$ (line 13), constructs a temporal window $[t_s^x,\, t_e^x + \Delta t]$ for each event (line 14), and increments a counter whenever a $Y$-event start time falls inside this window (lines 15-16), returning the hit count $h$.

The function \texttt{IndependentFrac} (lines 18-27) returns $0$ if service $Y$
has no anomalous events (line 19). Otherwise, it iterates over all $Y$-events
(line 21), checks whether each event is covered by any $X$-window
(lines 22-25), and counts uncovered events (line 26). It returns the uncovered
fraction $u / |\{t_s^y\}|$ (line 27), i.e., an empirical estimate of
$P(Y \mid \neg X)$.

The main loop iterates over all ordered service pairs $(X,Y)$ with $X \neq Y$
(line~28). Pairs belonging to the same application are skipped using
$\mathrm{AppOf}(X) = \mathrm{AppOf}(Y)$ (lines~29--31). The anomalous time
arrays $\{t_s^x, t_e^x, t_s^y, t_e^y\}$ are extracted from $D_a$ (line~32).
If either service has no anomalous events, the pair is skipped
(lines~33--35). The total number of anomalous events of $X$, $N_X$, is then
computed (line~36), followed by the hit count $h$ using \texttt{TemporalHits}
(line~37). The conditional probabilities(line 38-43) are estimated based on formulation explained in section 
Finally, the algorithm appends as a row in table $T$ (line~44). After all service pairs are processed, the completed table $T$ is returned (line~46).

\begin{algorithm}[!t]
\caption{Compute cross-application causal probabilities}
\label{alg:prob}
\begin{algorithmic}[1]
\REQUIRE
  Cleaned trace $\mathcal{D}$ with spans
  $(\text{spanID}, \text{traceID}, \text{service\_name}, \text{startTime}, \text{duration}, \text{error\_tag})$,
  application map $\text{AppOf}(\cdot)$, time window $\Delta t$, duration anomaly threshold $z$,
  minimum events per service $M$.
\ENSURE
  PN/PS/PNS table $\mathcal{T}$ over cross-application service pairs.

\STATE Load $\mathcal{D}$ from CSV.
\STATE For each span, parse $\text{startTime}$ into timestamp $t_s$.
\STATE Convert $\text{duration}$ to seconds and compute end time $t_e \gets t_s + \text{duration}$.
\STATE For each service $s$, compute mean $\mu_s$ and standard deviation $\sigma_s$ of span durations.
\STATE Set per-service duration threshold $\tau_s \gets \mu_s + z \cdot \sigma_s$.
\STATE Mark span $i$ as anomalous if $\text{error\_tag}_i = \text{True}$ \textbf{or}
       $\text{duration}_i > \tau_{\text{service\_name}_i}$.
\STATE Let $\mathcal{D}_a$ be the subset of anomalous spans.
\STATE For each service $s$, count anomalous spans $n_s$ in $\mathcal{D}_a$ and
       keep only services with $n_s \ge M$.
\STATE For each remaining service $s$, sort spans by $t_s$ and store arrays $t_s^s$ and $t_e^s$.
\STATE Initialize empty result table $\mathcal{T}$.

\STATE \textbf{function} $\text{TemporalHits}(t_s^X, t_e^X, t_s^Y, \Delta t)$
  \STATE \hspace{0.4cm} $h \gets 0$
  \STATE \hspace{0.4cm} \textbf{for} each event $i$ of $X$ \textbf{do}
    \STATE \hspace{0.8cm} window $\gets [t_{s,i}^X,\, t_{e,i}^X + \Delta t]$
    \STATE \hspace{0.8cm} \textbf{if} any $t_s^Y$ falls in window \textbf{then} $h \gets h + 1$
  \STATE \hspace{0.4cm} \textbf{return} $h$
\STATE \textbf{end function}

\STATE \textbf{function} $\text{IndependentFrac}(t_s^X, t_e^X, t_s^Y, \Delta t)$
  \STATE \hspace{0.4cm} \textbf{if} $|t_s^Y| = 0$ \textbf{then return} $0.0$
  \STATE \hspace{0.4cm} $u \gets 0$
  \STATE \hspace{0.4cm} \textbf{for} each $t \in t_s^Y$ \textbf{do}
    \STATE \hspace{0.8cm} covered $\gets$ \textbf{False}
    \STATE \hspace{0.8cm} \textbf{for} each event $i$ of $X$ \textbf{do}
      \STATE \hspace{1.2cm} window $\gets [t_{s,i}^X,\, t_{e,i}^X + \Delta t]$
      \STATE \hspace{1.2cm} \textbf{if} $t$ in window \textbf{then} covered $\gets$ True
    \STATE \hspace{0.8cm} \textbf{if} not covered \textbf{then} $u \gets u + 1$
  \STATE \hspace{0.4cm} \textbf{return} $u / |t_s^Y|$
\STATE \textbf{end function}

\FORALL{ordered pairs of services $(X,Y)$ with $X \neq Y$}
  \IF{$\text{AppOf}(X) = \text{AppOf}(Y)$}
     \STATE \textbf{continue} \COMMENT{skip same-application pairs}
  \ENDIF
  \STATE Extract $t_s^X, t_e^X, t_s^Y$ from $\mathcal{D}_a$.
  \IF{$|t_s^X| = 0$ \textbf{or} $|t_s^Y| = 0$}
     \STATE \textbf{continue}
  \ENDIF
  \STATE $N_X \gets |t_s^X|$
  \STATE $h \gets \text{TemporalHits}(t_s^X, t_e^X, t_s^Y, \Delta t)$
  \STATE $P(Y|X) \gets h / N_X$
  \STATE $P(Y|\neg X) \gets \text{IndependentFrac}(t_s^X, t_e^X, t_s^Y, \Delta t)$
  \STATE $PNS \gets \max(0,\, P(Y|X) - P(Y|\neg X))$
  \STATE $PN \gets \begin{cases}
      PNS / P(Y|X) & \text{if } P(Y|X) > 0 \\
      0 & \text{otherwise}
    \end{cases}$
  \STATE $PS \gets \begin{cases}
      PNS / \bigl(1 - P(Y|\neg X)\bigr) & \text{if } 1 - P(Y|\neg X) > 0 \\
      0 & \text{otherwise}
    \end{cases}$
  \STATE Append row $(X,Y,\text{AppOf}(X),\text{AppOf}(Y),]\newline P(Y|X),P(Y|\neg X),PNS,PN,PS,N_X,|t_s^Y|)$ to $\mathcal{T}$.
\ENDFOR
\STATE \textbf{return} $\mathcal{T}$.
\end{algorithmic}
\end{algorithm}

The allocation of microservices across worker nodes under a multi-objective optimization approach is formalized in \Cref{alg:spo}. It starts by importing all input data collected during load test, probability values generated by the method explained in\Cref{alg:prob}, Interference graphs(STIGs) generated by STIG simulator. It parses response time logs and computes a robust latency for each service using the median of p95 values (line 1). It then processes the PNS table, mapping each (X, Y) relation to canonical service keys for consistency across all artifacts (line 2). All STIG graphs are loaded and their interference weights aggregated, yielding one STIG based contention value per service pair (line 3). The global service set is built as the union of services appearing in logs, the PNS table, and the STIG (line 4). For each service pair present in both the causal table and STIG graphs, the algorithm computes a causal strength $\alpha(s,t)$ from PN, PS, and PNS (lines 5-7). Expected interference E(s,t) is then defined as $\alpha(s,t)$ times the STIG weight, capturing causal-topological co-location cost (line 8). In parallel, a communication edge set is built for all service pairs that exchange requests, to account for cross-node penalties (Line 9). Each service receives a priority score obtained by summing all incoming and outgoing E(s,t), identifying the most placement-critical services (lines 10-12). A multi-objective Optuna study is created, and each trial samples a candidate node placement for all services  (lines 13-27). For each trial, total interference is computed by summing E(s,t) over co-located pairs, while latency is adjusted with penalties for cross-node communication. After all trials, Pareto-optimal placements are extracted, one is selected (preferring lower interference), and the algorithm returns the chosen placement, its objectives, and the priority scores (lines 28-30).

\begin{algorithm}[!t]
\caption{Service Placement by Optimizer (SPO)}
\label{alg:spo}
\begin{algorithmic}[1]
\REQUIRE
    Probability table (PS,PN,PNS) $\mathcal{P}$ (columns $X,Y,\text{App}(X),\text{App}(Y),PN,PS,PNS$),
    response-time log $\mathcal{R}$,
    STIG graphs directory $\mathcal{G}$,
    set of nodes $\mathcal{N}$,
    number of optimization trials $T$.
\ENSURE
    Best placement $\pi^\star : \mathcal{S} \rightarrow \mathcal{N}$,
    total interference $I(\pi^\star)$,
    total latency $L(\pi^\star)$,
    service priorities $\text{Priority}(s)$.
\STATE Parse $\mathcal{R}$ and compute median $p95$ latency $L(s)$ for each service $s$.
\STATE Parse $\mathcal{P}$ and map $(X,Y,\text{App}(X),\text{App}(Y))$
       to canonical service keys $(s,t)$ using the service-name map.
\STATE Parse all STIG graphs in $\mathcal{G}$ and aggregate
       interference weights $w_{\text{STIG}}(s,t)$ for each service pair $(s,t)$.
\STATE Let $\mathcal{S}$ be the union of all services that appear
       in $\mathcal{R}$, $\mathcal{P}$ or $\mathcal{G}$.
\STATE For every pair $(s,t)$ that appears in both $\mathcal{P}$ and $\mathcal{G}$:
       \STATE \hspace{0.75cm} read $PN(s,t)$, $PS(s,t)$, $PNS(s,t)$ from $\mathcal{P}$;
       \STATE \hspace{0.75cm} compute causal factor
           $\alpha(s,t) \gets PN(s,t) + PS(s,t) - PNS(s,t)$;
       \STATE \hspace{0.75cm} compute expected interference
           $E(s,t) \gets \alpha(s,t) \cdot w_{\text{STIG}}(s,t)$.
\STATE Build communication edge set $\mathcal{C}$ as undirected pairs
       $\{s,t\}$ whenever $(s,t)$ or $(t,s)$ exists in $\mathcal{P}$.
\FORALL{services $\{s\} \in \mathcal{S}$}
    \STATE $\text{Priority}(s) \gets
      \sum_{t} E(s,t)$  
\ENDFOR

\FOR{$k = 1$ \TO $T$}
    \STATE Optimizer proposes a placement
           $\pi_k : \mathcal{S} \rightarrow \mathcal{N}$
           by sampling a node index for each service.
    \STATE $I(\pi_k) \gets 0$; \quad $L(\pi_k) \gets \sum_{s \in \mathcal{S}} L(s)$.
    \FORALL{ordered pairs $(s,t)$ with defined $E(s,t)$}
        \IF{$\pi_k(s) = \pi_k(t)$}
            \STATE $I(\pi_k) \gets I(\pi_k) + E(s,t)$.
        \ENDIF
    \ENDFOR
    \FORALL{edges $\{s,t\} \in \mathcal{C}$}
        \IF{$\pi_k(s) \neq \pi_k(t)$}
            \STATE $L(\pi_k) \gets L(\pi_k) + \text{NET\_PENALTY\_MS}$.
        \ENDIF
    \ENDFOR
    \STATE Report objective vector
        $(I(\pi_k), L(\pi_k))$ to Optimizer.
\ENDFOR

\STATE Retrieve the Pareto-optimal trials from the Optimizer study.
\STATE Select $\pi^\star$ from the Pareto front using a tie-breaking rule
       (e.g., minimum $I(\pi)$, then minimum $L(\pi)$).
\STATE \textbf{return} $\pi^\star$, $I(\pi^\star)$, $L(\pi^\star)$,
       and $\text{Priority}(s)$ for all $s$.
\end{algorithmic}
\end{algorithm}

\section{Evaluation}\label{sec:evaluation}
\subsection{Experimental Setup}\label{sec:Setup}
\begin{figure}[htbp]
\centerline{\includegraphics[width=0.5\textwidth]{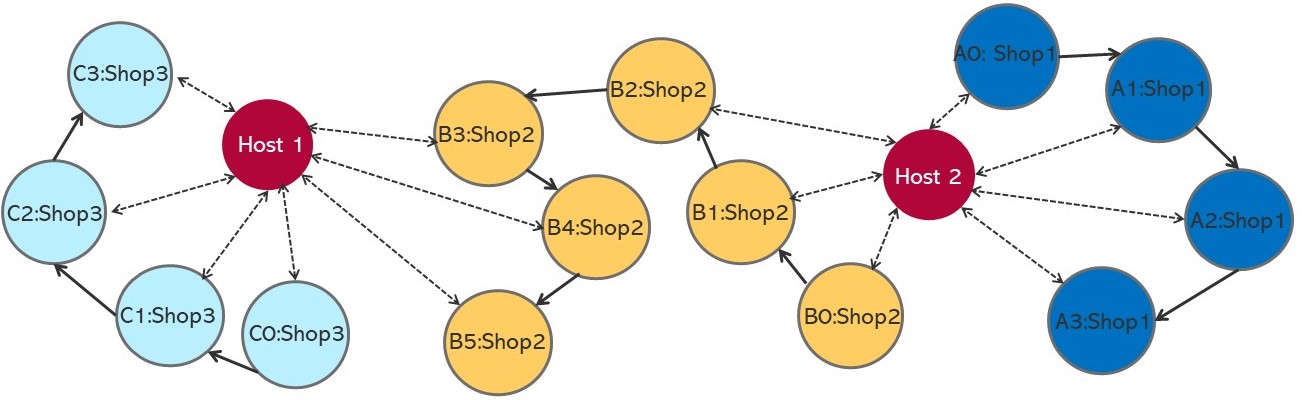}}
\caption{knowledge graph}
\label{fig:Kg}
\end{figure}

\textbf{Testbed Infrastructure} -- In our in-house developed testbed, all experiments ran on a single Ubuntu VM (can add more VMs as per complexity of architecture), provisioned with 32 vCPUs, 50 GiB RAM, and 200 GiB SSD storage. We deployed a Kind Kubernetes cluster with one control plane node and two worker nodes. We deploy three microservice applications (Sock-Shop\footnote{SockShop: https://microservices-demo.github.io/}, TeaStore\footnote{Tea-Store: https://github.com/DaGeRe/TeaStore}, and Bookinfo\footnote{Bookinfo: https://istio.io/latest/docs/examples/bookinfo/}) across various Kubernetes namespaces. Three HTTPRoutes (/sock, /tea, /book) received traffic that was entered over an Istio Gateway (Gateway API). Every application namespace had sidecar injection enabled. There was no inter-VM traffic since the cluster was using the default type CNI. Prometheus (scrape interval: 15s) gathered metrics and surfaced them in Kiali. Jaeger (all-in-one) collected distributed traces using an Istio Telemetry resource that allowed tracing. Logs were recorded using the stdout/stderr container.
A detailed knowledge graph (\Cref{fig:Kg}) shows three deployed shop applications on a cluster. 


\textbf{Reproducibility} -- To create the type cluster, deploy Istio, and execute all experiments, we offer k6 scripts, Kubernetes manifests (Namespaces, Gateway/HTTPRoutes, DestinationRules), and a run script\footnote{See cluster script at https://doi.org/10.5281/zenodo.18900711}. Each run used a different set of random seeds and scenario ordering. \newline
To demonstrate the resilience of our proposed method and its benefits over several state-of-the-art approaches, we carry out intensive experiments. Our collected datasets consist of specific of incidents mentioned below, collected from multi node Kubernetes cluster. We created scenario-based tests that represent actual operational difficulties in a multi-node cluster where interference happened. The selected scenario describes adaptation goals commonly found in microservices applications. We quantify changes in system-level parameters, including error rate (service failure, "req\_failed"), interference score, and reaction time, before and after adaptation in order to assess adaptation performance across all circumstances. \newline
\textbf{Test Scenario} -- We simulated a surge in resource demand to mimic an abrupt jump in resource demand, similar to a flash sale or unanticipated traffic spike. As the load rises, we monitor how interference levels and response rates vary. This maps to our three research questions (RQ.1, RQ.2, RQ.3) and a set of specified metrics (\Cref{tab:metrics}) computed (sampled) by the monitoring stack. Next we detail the directives we followed to simulate the  surge scenario. 
\begin{itemize}
    \item Generate traffic spike using k6, increasing CPU, memory, and request load across services.
    \item Check whether the system can stop co-location when under stress.
    \item Monitor response degradation and re-optimization frequency.
    \item Analyze Optimizer stability and MAPE-K loop responsiveness as system pressure rises.
\end{itemize}

\textbf{Metrics to Collect per Scenario} -- We collect metrics (\Cref{tab:metrics}) that characterize system behavior during each test. The optimizer logs planning-side signals (placement, interference\_score, response\_score, iteration, migration count, convergence time), while Prometheus/Istio provide runtime measures including end-to-end response time (RT), tail latency(\(L^{p95}\)/\(L^{p99}\) latency), error rate \(\varepsilon_t\) and node resource utilization. Response time is considered primary metric at macro level as it encompasses full end-to-end experience in test scenario. We monitor, compare, and display response times across time, services, or nodes since they are continuous and timestamped. When assessing service placement under interference, it is a more accurate statistic since it directly correlates to user-perceived delay. Span duration is a secondary metric at micro level which is a time required for a particular microservice or function to complete a trace. It is also important when we want to analyze service performance bottlenecks, also calculate the internal delay for each microservice (payment service vs. cart service in SockShop , for example) and examine breakdown of total time per component. We can measure any interference and its mitigation after re-deployment (RQ.1, RQ.2) and sensitivity to missing knowledge (RQ.3), by correlating these streams (SLO breach events → re-optimization latency → new placement → change in interference and latency). We can also demonstrate that the causal, STIG-guided planner maintains interference magnitude reduction across applications (RQ.1).

\begin{table}[!t]
\caption{Monitoring Metrics and Logs}
\vspace{-6pt}
\centering
\footnotesize
\setlength{\tabcolsep}{3pt}
\renewcommand{\arraystretch}{0.9}
\begin{tabular}{|p{0.33\columnwidth}|p{0.23\columnwidth}|p{0.34\columnwidth}|}
\hline
\textbf{Metric} & \textbf{Tool} & \textbf{Purpose} \\
\hline
Response time & Prometheus, Istio & SLO violations, user experience \\
\hline
\texttt{request duration} (P95) & Prometheus & Latency degradation \\
\hline
Placement logs (.json) & Test script & Service-to-node mapping \\
\hline
SLO violations & Prometheus, Istio, Jaeger & Responsiveness \\
\hline
Node utilization & \texttt{kubectl top node}, metrics-server & Affected nodes \\
\hline
\end{tabular}
\label{tab:metrics}
\end{table}

\subsection{Scenario Based Analysis}
\textbf{Surge in resource demand} --
To examine how the multi-node cluster responds to sudden increases in resource demand, we use a load generator to send traffic through the Istio ingress and monitor the cluster’s behavior during spikes. We perform load tests at regular intervals. In each test run, we first establish a baseline using Kubernetes’ default placement for all microservices, then run a k6 workload while collecting distributed traces and performance metrics. These default-cluster multiple test run measure the extent of cross-service interference before any adaptive placement strategy is applied. \Cref{tab:baseline_load} shows a clear performance imbalance among the services across the four test runs. 

\begin{table}[htbp]
\caption{Default Cluster (without optimization) across multiple test runs, p95 in milliseconds ms}
\label{tab:baseline_load}
\vspace{-6pt}
\centering
\footnotesize
\setlength{\tabcolsep}{4pt}
\renewcommand{\arraystretch}{1.12}
\begin{tabular}{|p{0.10\columnwidth}|p{0.16\columnwidth}|p{0.17\columnwidth}|p{0.15\columnwidth}|p{0.20\columnwidth}|}
\hline
\textbf{Run} & \textbf{p95 (book)} & \textbf{p95 (sock)} & \textbf{p95 (tea)} & \textbf{req\_failed (\%)} \\
\hline
1 & 10250 & 47.77 & 202.24 & 24.62 \\
\hline
2 & 10340 & 47.71 & 177.11 & 24.75 \\
\hline
3 & 10300 & 47.98 & 143.41 & 24.59 \\
\hline
4 & 10290 & 47.93 & 143.96 & 24.57 \\
\hline
\textbf{Mean} & \textbf{10295.00} & \textbf{47.85} & \textbf{166.68} & \textbf{24.63} \\
\hline
\textbf{Std}  & \textbf{32.02} & \textbf{0.11} & \textbf{24.65} & \textbf{0.07} \\
\hline
\end{tabular}
\end{table}

We collect traces after load test and analysis each trace, its start time, duration, latency and failure status. During the test window, we collected 24000 spans. For example, in Test-1 (\Cref{tab:baseline_load}), 5,940 spans were recorded, 414 of which were identified as error spans. \Cref{tab:top_services_p95} contains the latency results obtained on the collected trace.
\begin{table}[!t]
  \caption{Top services by  latency (P50, 95, 99, Max in milliseconds ms)}
  \label{tab:top_services_p95}
  \vspace{-6pt}
  \centering
  \footnotesize
  \begin{tabular}{|p{0.27\columnwidth}|p{0.20\columnwidth}|p{0.05\columnwidth}|p{0.05\columnwidth}|p{0.05\columnwidth}|p{0.05\columnwidth}|}

    \hline
    \textbf{Service} & \textbf{Fail rate (\%)} & \textbf{P50} & \textbf{P95} & \textbf{P99} & \textbf{Max} \\
    \hline
    productpage.bookinfo    & 0.00  & 21.63 & 64.95 & 67.61 & 82.52 \\
    \hline
    details.bookinfo        & 0.00  &  2.00 & 42.60 & 43.63 & 46.90 \\
    \hline
    front-end.sock-shop     & 0.00  &  1.75 & 42.33 & 42.81 & 43.13 \\
    \hline
    reviews.bookinfo        & 0.00  &  4.61 &  6.03 &  7.25 &  7.67 \\
    \hline
    webui.teastore & 50.00 &  2.37 &  3.49 &  4.16 &  9.63 \\
    \hline
    ratings.bookinfo        & 0.00  &  0.86 &  1.07 &  1.18 &  1.26 \\
    \hline
  \end{tabular}
\end{table}

\begin{figure}[t]
\centerline{\includegraphics[width=0.43\textwidth,keepaspectratio]{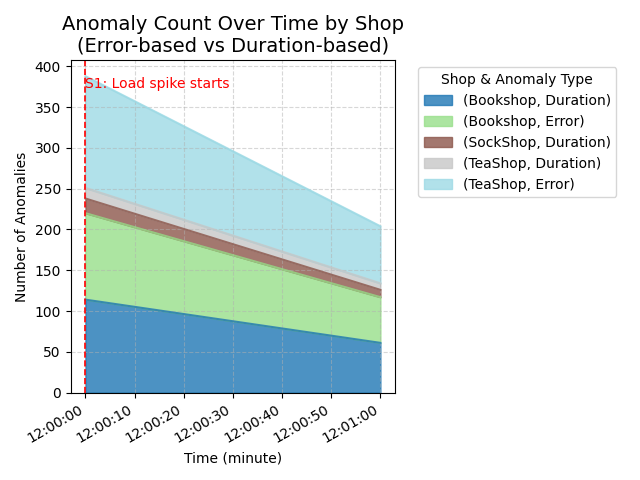}}
\caption{Interference Anomaly (Service Failure/Error) vs Request Duration}
\label{fig:anomlyplot}
\end{figure}

This workload produced an overall failure rate of 7\%. Tail latency was analyzed using the 95th percentile (p95) to highlight services that struggle under peak load. The main hotspot is productpage.bookinfo. Details.bookinfo and front-end.sock-shop also show elevated p95 values. In contrast, reviews.bookinfo and ratings.bookinfo remain fast and stable (p95 = 6 ms and 1 ms). webui.teastore service is unusual, it has a very high error rate (approx. 50\%) but low p95 latency (approx. 3–4 ms), indicating many quick failures (e.g., misrouting, unavailable backends, or quota issues) rather than slow responses. Overall, the data match the architecture. productpage service acts as a gateway and accumulates tail latency. Sock-Shop’s front-end shows milder stress, TeaStore appears misconfigured or under-provisioned. This span-level view identifies where to focus optimization such as scaling, timeout tuning, and resilience improvements.
The broader goal is to detect interference anomalies when multiple shops run concurrently on the same cluster. \Cref{fig:anomlyplot} shows the minute-level distribution of anomalies during \textit{Load Spike} scenario across shops. The red dashed line marks the onset of the spike. Bookshop and TeaStore exhibit far more anomalies than Sock-Shop just after concurrent start-up. TeaStore is dominated by error-based anomalies (instability/failures under load), while Bookshop shows many duration-based anomalies (queuing and slowdown from resource contention). Anomalies gradually decline over time, consistent with cluster adaptation (e.g., throttling or queue stabilization). SockShop remains largely unaffected, suggesting lower resource demand or indirect isolation. These patterns support the hypothesis that co-located services under concurrent traffic experience interference, producing distinct anomaly profiles and revealing imbalanced resource contention on the shared infrastructure.

 Next, we process these traces to compute the causal interference measures PN, PS, and PNS for all observed pairs of services (see \Cref{alg:prob}).
 
\textbf{Cluster Reconfiguration Plan} -- In the reconfiguration plan, the services are ranked according to their likelihood of causing the interference anomaly. Since interference occurs in both directions, the strategy may identify abnormal services on each side of an interference relationship as the source and the target. When an anomalous service interferes with services from other applications, our method ranks services by the likelihood that they are necessary and sufficient, explained in \Cref{causility-STIG}. We derived the probabilities displayed in \Cref{tab:causal_interference} after applying these formulas to the interference anomaly data.
\begin{table}[!t]
  \caption{Causal interference probabilities between services.}
  \label{tab:causal_interference}
  \vspace{-6pt}
  \centering
  \footnotesize
  \setlength{\tabcolsep}{2pt}
  \renewcommand{\arraystretch}{1.0}
  \begin{tabular}{|l|l|c|c|c|}
    \hline
    \textbf{X (source)} & \textbf{Y (target)} & \textbf{PNS} & \textbf{PN} & \textbf{PS} \\
    \hline
    webui.teastore & reviews.bookinfo &   99.79 & 100.00 &  99.79 \\
    \hline
   webui.teastore & ratings.bookinfo     &  96.03 & 100.00 &  96.03 \\
    \hline
    webui.teastore & productpage.bookinfo &  42.68 & 100.00 &  42.68 \\
    \hline
    productpage.bookinfo    & webui.teastore & 92.68 &  92.68 & 100.00 \\
   \hline
    ratings.bookinfo        & webui.teastore &  81.12 &  81.12 & 100.00 \\
    \hline
    reviews.bookinfo        &  webui.teastore &  81.12 &  81.12 & 100.00 \\
    \hline
  \end{tabular}
\end{table}
\newline
\textbf{Service Placement} --
The placement strategy, explained in \Cref{algo}, jointly considers cross-application service interference and deployment-induced latency. Based on STIG interference weights and causal PN/PS/PNS estimates, the optimiser explores placements that minimise expected interference while constraining communication overhead. The resulting configurations indicate which services should be co-located and which should be separated across nodes to reduce contention, thereby guiding reconfiguration decisions that enhance performance stability in multi-tenant deployments.
\Cref{tab:new_config} shows a comparison between default and best possible placement setting suggested by Optuna. 
\begin{table}[!t]
  \caption{Optuna suggested configuration (Best Possible)}
  \label{tab:new_config}
  \vspace{-6pt}
  \centering
  \footnotesize
  \setlength{\tabcolsep}{4pt}
  \renewcommand{\arraystretch}{1.0}
 \begin{tabular}{|l|l|l|}
    \hline
    \textbf{Service} & \textbf{Default placement} & \textbf{Optimized placement} \\
    \hline
    productpage.bookinfo & kind-cluster-worker2 & kind-cluster-worker1 \\
    \hline
    ratings.bookinfo & kind-cluster-worker1 & kind-cluster-worker2 \\
    \hline
    reviews.bookinfo & kind-cluster-worker1 & kind-cluster-worker2 \\
    \hline
    front-end.sock-shop & kind-cluster-worker1 & kind-cluster-worker2 \\
    \hline
    webui.teastore & kind-cluster-worker1 & kind-cluster-worker1 \\
    \hline
  \end{tabular}
  \vspace{2pt}
\end{table}

For instance, under certain workload, the Optuna-based optimizer identifies webui.teastore and productpage.bookinfo as the most interference-sensitive services, followed by the front-end and two additional Bookinfo backend services. Using these priorities together with the learned interference matrix, it places Teastore’s web UI alongside a small group of low-priority services on kind-cluster-worker1, while placing most Bookinfo and Sock-shop services on kind-cluster-worker2. This placement cleanly separates all highly interfering service pairs, achieving a total interference score of 45.67 and keeping the overall end-to-end latency around $2.3 \times 10^{6} \,\text{ms}$. 

\begin{figure}[t]
\centerline{\includegraphics[width=0.45\textwidth, height=0.25\textheight]{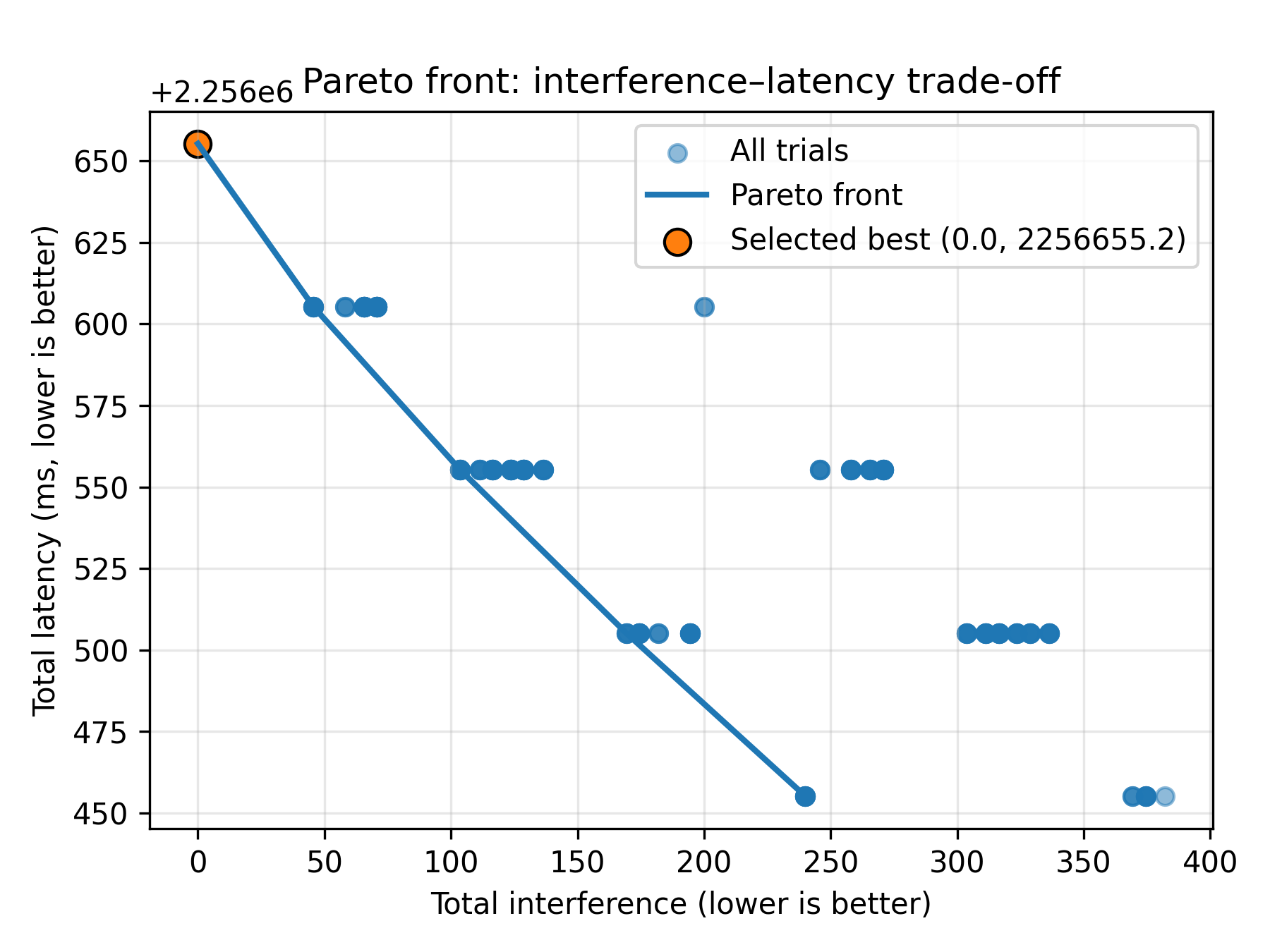}}
\caption{Trade-off between total expected interference and total latency}
\label{fig:pareto_plot}
\end{figure}

The Fig.~\ref{fig:pareto_plot}  shows the trade-off between total expected interference and total latency for all service placements evaluated by Optuna. Each blue dot is a candidate placement, scored on two objectives: (1) total interference (i.e., the sum of expected interference over all service pairs derived from STIG weights and PN/PS/PNS metrics) and (2) total latency (i.e., the overall response-time cost including cross-node communication penalties). Reducing interference generally spreads services across nodes (raising latency), while reducing latency tends to co-locate them (raising interference), so the optimization produces a trade-off between them.

The solid blue line is the Pareto front configurations where any further reduction in interference would increase latency and vice versa. The orange point marks the selected placement, with the adjacent values giving its interference–latency pair. It lies on or near the Pareto front, indicating an efficient compromise. Latency values are plotted with a constant vertical offset ($2.256 \times 10^{6}$ subtracted) so that the axis focuses on meaningful variation rather than absolute magnitudes. Overall, we show that Optuna explores a wide range of placements and that the chosen configuration is a trade-off between performance and interference mitigation in the multi-tenant cluster.
\newline
\textbf{Re-Deployment Results} -- 
In this section, we explain the service placement results for the optimized cluster suggested by our Optuna placement method. \Cref{tab:opt_cp1_load_tests} reports the raw p95 latency and error statistics for the optimized placement across four test runs for the suggested placement. Book-info and Sock-shop remain in the same general latency regime as in the baseline -- book stays around 10.1 s p95 and sock around 47 ms, both with very small variance, which indicates that the optimization does not change their worst-case behaviour. Conversely, TeaStore shows a lower p95 latency of about 8.8 ms, suggesting an improvement over the default cluster results in the previous \Cref{tab:baseline_load}. The HTTP failure rate (req\_failed) under the optimized placement is consistently around 32.7 $\%$ (slightly higher than in the default configuration).
\begin{table}[!t]
  \caption{Optimized cluster-placement- load test run}
  \label{tab:opt_cp1_load_tests}
  \vspace{-6pt}
  \centering
  \footnotesize
  \setlength{\tabcolsep}{1pt}
  \renewcommand{\arraystretch}{1.0}
  \begin{tabular}{|l|r|r|r|r|}
    \hline
    \textbf{Run} & \textbf{p95 (book, ms)} & \textbf{p95 (sock, ms)} & \textbf{p95 (tea, ms)} & \textbf{req\_failed (\%)} \\
    \hline
    1   & 10\,100.00 & 47.24 & 8.80 & 32.58 \\
    \hline
    2   & 10\,090.00 & 47.25 & 8.74 & 32.63 \\
    \hline
    3   & 10\,190.00 & 47.36 & 8.76 & 32.71 \\
    \hline
    4   & 10\,210.00 & 47.45 & 8.83 & 32.83 \\
    \hline
    \textbf{Mean} & \textbf{10\,147.50} & \textbf{47.32} & \textbf{8.78} & \textbf{32.69} \\
    \hline
    Std & 61.31 & 0.10 & 0.04 & 0.11 \\
    \hline
  \end{tabular}
\end{table}

\Cref{tab:default_vs_optimized_combined} now combines default and optimized cluster statistics (mean $\pm$ standard deviation) with the absolute and relative
$\Delta$. Book-info and Sock-shop present modest p95 latency reductions of about $1$--$1.5\%$, showing that the optimized
placement preserves their performance while giving small gains. TeaStore improves dramatically: its p95 latency drops from
$166.68 \pm 24.65$ ms to $8.78 \pm 0.04$ ms, corresponding to $\Delta_{\text{abs}} = -157.90$ ms and $\Delta\% = -94.73$,
which indicates that a major bottleneck or interference source has been removed. At the same time, the mean HTTP failure rate
increases from $24.63 \pm 0.07\%$ to $32.69 \pm 0.11\%$ ($\Delta_{\text{abs}} = +8.06$ percentage points,
$\Delta\% = +32.72$). Overall, the optimized placement achieves strong latency reductions, especially for TeaStore, but at the cost
of an increase in request failures.
\begin{table}[!t]
  \caption{Default vs optimized cluster. p95 in ms; failure in \%.}
  \label{tab:default_vs_optimized_combined}
  \vspace{-6pt}
  \centering
  \scriptsize
  \setlength{\tabcolsep}{1pt}
  \renewcommand{\arraystretch}{0.9}
  \begin{tabular}{|l|r|r|r|r|}
    \hline
    \textbf{Metric} &
    \textbf{Default ($\mu\!\pm\!\sigma$)} &
    \textbf{Optimized ($\mu\!\pm\!\sigma$)} &
    \textbf{$\Delta$ abs} &
    \textbf{$\Delta$ \%} \\
    \hline
    p95 (book, ms) &
    $10\,295.00 \pm 32.02$ &
    $10\,147.50 \pm 61.31$ &
    $-147.50$ &
    $-1.43$ \\
    \hline
    p95 (sock, ms) &
    $47.85 \pm 0.11$ &
    $47.32 \pm 0.10$ &
    $-0.53$ &
    $-1.11$ \\
    \hline
    p95 (tea, ms) &
    $166.68 \pm 24.65$ &
    $8.78 \pm 0.04$ &
    $-157.90$ &
    $-94.73$ \\
    \hline
    req\_failed (\%) &
    $24.63 \pm 0.07$ &
    $32.69 \pm 0.11$ &
    $+8.06$ &
    $+32.72$ \\
    \hline
  \end{tabular}
\end{table}

\subsection{Comparison with Baseline:} 
We selected few baselines intentionally to compare with selected Optuna optimizer for service placement problem in our study. They reflect standard practice for placement without access to causal information, and all are evaluated under the same search budget as Optuna. The Optuna (no causal, latency-only) configuration is an ablation of our proposed method. It adopts the same optimizer and search space as the full causal Optuna variant, but we remove PN/PS/PNS and STIG from the objective function. This design isolates the contribution of causal information by holding all other components fixed.
The Genetic Algorithms (GA)\cite{saravanan2002multi} and Reinforcement Learning (RL) \cite{shakya2023reinforcement} baselines are widely used “state-of-the-art” meta-heuristics for placement and scheduling problems. In our setup (\Cref{tab:baseline_comp}), both optimize exactly the same latency-only objective as Optuna (i.e., non-causal). 
\begin{table}[!t]
  \caption{Comparison of optimization strategies.}
  \label{tab:baseline_comp}
  \vspace{-6pt}
  \centering
  \scriptsize
  \setlength{\tabcolsep}{1pt}
  \renewcommand{\arraystretch}{0.9}
  \begin{tabular}{|l|r|r|}
    \hline
    \textbf{Method} &
    \textbf{Interference} &
    \textbf{Latency (ms)} \\
    \hline
    Optuna (full causal) &
    $997.30$ &
    $7\,474\,879.01$ \\
    \hline
    Optuna (no causal, lat-only) &
    $1\,763.49$ &
    $7\,409\,925.13$ \\
    \hline
    GA baseline &
    $1\,763.49$ &
    $7\,409\,925.13$ \\
    \hline
    RL baseline &
    $1\,763.49$ &
    $7\,409\,925.13$ \\
    \hline
  \end{tabular}
\end{table}

This allows us to distinguish whether any observed improvements are due to the optimization algorithm (Optuna vs GA/RL) or due to the objective formulation (causal vs non-causal). All four strategies are evaluated using the same objective, full-causal interference (PN/PS/PNS, weighted by STIG) combined with placement-sensitive latency. We observe that all latency-only strategies (Optuna no-causal, GA, RL) converge to the same placement. They achieve essentially identical interference and latency. This is consistent with the fact that they optimize the same latency-only objective, and it also indicates that Optuna is competitive with GA and RL as a generic black-box optimizer. In contrast, the full-causal Optuna configuration converges to a different placement, the interference decreases from approximately 1763 to 997 (about a 43\% reduction), at the cost of a small increase in latency (around +0.9\%). This baseline design supports the following interpretation, relative to realistic “black-box” heuristics (GA/RL) that only observe latency, augmenting the objective with PN/PS/PNS and STIG systematically steers Optuna toward placements with substantially lower causal interference. Since GA and RL reproduce the latency-only Optuna solution, the observed improvements cannot be attributed merely to Optuna being a stronger optimizer, they arise specifically from incorporating causal knowledge into the placement objective.

\subsection{Answers to Research Questions}
\textbf{RQ.1 Answer:} \textbf{Yes}, interference is high both in terms of density (many contending services per node) and impact (large latency inflation). Under the default cluster placement, interference appears in how far it spreads across nodes and how much it worsens latency. Services from BookInfo, Sock-shop, and TeaStore are densely co-located on the same Kubernetes worker node, so they compete for CPU, memory, and network queues. This amplifies even moderate cross-traffic and produces clear contention in the traces (\Cref{fig:anomlyplot}). The PN, PS, and PNS estimates (\Cref{tab:causal_interference}) capture this effect, for example, webui.teastore → reviews.bookinfo and webui.teastore → ratings.bookinfo both have PN = 100 with PNS/PS above 96, while productpage.bookinfo → webui.teastore reaches PNS = 92.68 and PS = 100 (inclusive the reverse edge also has PN = 92.68), all indicating strong causal interference when these services share a node. Because high-traffic services are concentrated on a small subset of nodes, the interference is localized but intense, driving p95 latencies from hundreds up to tens of thousands of milliseconds for some services that failed to serve the request. For instance,  webui.teastore service has failure rate 50\% larger than other services (\Cref{tab:top_services_p95}). Therefore, the higher density (contending services per node) and impact (latency) motivate a causal-aware placement as a means to mitigate interference and recover performance.\newline
\textbf{RQ.2 Answer}: \textbf{Yes}. Across all test runs, the causal Optuna model best recovers from performance degradation, shown in \Cref{tab:default_vs_optimized_combined}. Even as the number of services and communication patterns grows, the optimizer consistently finds placements with much lower interference than all non-causal baselines.
The results also expose a key trade-off: reducing interference usually means spreading services across nodes, but spreading them increases cross-node latency. The causal model learns placements that balance these forces, while non-causal methods either (a) over-concentrate services and cause interference, or (b) overspread them and inflate latency. Overall, the approach remains effective, minimizing both residual interference and the number of impacted nodes as connectivity complexity increases. \newline
\textbf{RQ.3 Answer:} \textbf{Yes}, causal knowledge (PNS and STIGs) has a discernible impact on placement effectiveness. The ablation experiments show that removing causal components degrades performance, as follows:
(a) Without PN/PS/PNS, the optimizer loses direction and strength of interference, leading to near-random co-locations and much higher total interference; (b) Without STIG (temporal–topological severity), it cannot distinguish mild from severe interference patterns, leading to suboptimal placements even when PN/PS/PNS are present; and (c) Removing both reduces the system to a latency-only optimizer, which consistently underperforms the causal model.
\Cref{tab:q3ans} shows an ablation study that covers a full causal model (PN/PS/PNS × STIGs knowledge) achieves zero interference with slightly less latency comparable to all ablations. With STIG only, interference rises slightly (2.4) with virtually no latency benefit. With PN/PS/PNS only, interference jumps sharply (300) despite similar latency. The latency-only baseline also finds an undetectable interference placement in this small case. However, without modeling interference structure, it is unlikely that its identification will be reliable under tighter constraints or higher contention. Each causal component adds distinct, non-redundant information, and their combination for near-optimal interference-aware placement.
\begin{table}[!t]
  \caption{Ablation summary (best possible values). All latency values in milliseconds, Interference is a score.}
  \label{tab:q3ans}
  \vspace{-6pt}
  \centering
  \footnotesize
  \setlength{\tabcolsep}{4pt}
  \renewcommand{\arraystretch}{1.0}
  \begin{tabular}{|l|c|c|}
    \hline
    \textbf{Configuration} & \textbf{Interference} & \textbf{Latency (ms)} \\
    \hline
    Causal (PN/PS/PNS $\times$ STIG)           & 0.000   & 1781622.249 \\
    \hline
    No PN/PS/PNS (STIG only)                   & 2.400   & 2256455.232 \\
    \hline
    No STIG (PN/PS/PNS only)                   & 300.000 & 2256505.232 \\
    \hline
    Latency-only (non-causal baseline)         & undetectable   & 2256405.232 \\
    \hline
  \end{tabular}
\end{table}

\section{Discussion}\label{sec:discussion}
\subsection{Implications} This work shows that treating service placement as a cross-application interference-aware, multi-objective optimization problem yields measurable gains in tail latency and stability for multi-application clusters. It indicates that topology-agnostic scheduling leaves performance on the table, even without changing code or resource limits. Placements guided by empirical interference profiles can significantly reduce p95 latency, especially for highly contended services like TeaStore. The framework exposes explicit trade-offs among latency, interference, and cross-node communication, making it useful as a decision-support tool for operators choosing Pareto-optimal placements that reflect current priorities (for example, isolation for critical services versus lower network cost at peak). More broadly, the results suggest that interference can inform service placement alongside CPU and memory usage. While we showed that in a reactive feedback control loop (MAPE-K), our causal approach could also inform self-stabilizing orchestration~\cite{cai2024self} and, eventually, be integrated with Kubernetes schedulers and autoscalers~\cite{incerto2023muopt,quattrocchi2024autoscaling,liang2026autoscaling}.
\subsection{Threats to validity} 
\textbf{Internal Validity} -- The interference model is learned from offline profiling with Optuna rather than being deployed in a live scheduler. As a result, it may not adapt to short-term workload shifts or concept drift in real deployments. Moreover, the estimation of “causal interference” relies on correlations that may be biased by unobserved confounding factors, potentially affecting the accuracy of the inferred relationships.

\textbf{External Validity} -- The evaluation relies on a fixed set of multi-application workloads and microservice topologies (Book, Sock, and Tea e-commerce stacks). These testbeds may not capture the diversity of interference patterns present in larger, heterogeneous clusters, or under conditions such as bursty traffic, GPU-intensive, or batch workloads. Consequently, the generalizability of the results to broader production contexts remains limited.

\textbf{Construct Validity} -- The analysis focuses on p95 response time and HTTP failure rate as primary quality-of-service (QoS) metrics. Other important dimensions, such as p99 latency, jitter, throughput, energy efficiency, or cost, are not optimized, and these may exhibit different behaviors. Our metric selection could restrict how comprehensively the study captures system performance.

\section{Conclusion \& Future Work}\label{sec:conclusion}
We introduced an interference-aware, cross-application placement method for containerized microservices in shared clusters. Our approach is driven by causal knowledge of counterfactual probabilities of sufficiency and necessity computed through a Spatio-Temporal representation model (STIG) of the interference phenomenon. By combining cross-application profiling, per-service latency baselines, and the STIG-based estimates of probabilities of sufficiency and necessity (PN/PS/PNS), we formulated service placement as a multi-objective optimization problem that minimizes interference and improves response times under network and isolation constraints. Using Optuna’s multi-objective engine, our study explored trade-offs between latency, interference, and communication cost to derive the Pareto-optimal placements that reflect operator preferences. The results indicate that a causality-driven multi-objective formulation provides operators with explicit control over interference, latency, and communication overhead. In future work we plan to integrate the optimizer with an online scheduler, extend the interference model with additional QoS and reliability metrics, and scale the experiments to larger and more heterogeneous clusters.
\clearpage 
\bibliographystyle{IEEEtran}
\bibliography{references}

\end{document}